\documentclass[preprint,floatfix]{revtex4-1}

\usepackage{amsmath}
\usepackage{graphicx}

\newcommand{\bra}[1]{\langle #1 |}
\newcommand{\ket}[1]{| #1 \rangle}

\begin{document}

\title{Why Quantum Coherence is Not Important in the Fenna-Matthews-Olsen Complex}

\author{David M. Wilkins}
\email{david.wilkins@chem.ox.ac.uk}
\affiliation{Physical And Theoretical Chemistry Laboratory, Oxford University, South Parks Road, Oxford, OX1 3QZ}

\author{Nikesh S. Dattani}
\email{dattani.nike@gmail.com}
\affiliation{Quantum Chemistry Laboratory, Department of Chemistry, Kyoto University, 606-8502, Kyoto, Japan}
\affiliation{Physical And Theoretical Chemistry Laboratory, Oxford University, South Parks Road, Oxford, OX1 3QZ}


\begin{abstract}
We develop and present an improvement to the conventional technique for solving the Hierarchical Equations of Motion which reduces the memory cost by more than 75\% while retaining the same convergence rate and accuracy. This allows for a full calculation of the population dynamics of the 24-site FMO trimer for long timescales with very little effort, and we present the first fully converged, exact results for the 7-site subsystem of the monomer, and for the full 24-site trimer. Owing to this new approach, our numerically exact 24-site, 2-exponential results are the most demanding HEOM calculations performed to date, surpassing the 50-site, 1-exponential results of Strumpfer and Schulten [2012, J. Chem. Thy. \& Comp., 8, 2808]. We then show where our exact 7-site results deviate from the approximation of Ishizaki and Fleming [2009, Proc. Natl. Acad. Sci. USA, 106, 17255]. Our exact results are then compared to calculations using the incoherent F\"orster theory, and it is found that the energy transfer from the antenna to the reaction centre occurs more than 50 times faster than the fluorescence lifetime of the excitation, whether or not coherence is considered. This means that coherence is not likely to improve the efficiency of the photosynthesis. In fact, the incoherent theory often tends to over-predict the rates of energy transfer, suggesting that in some cases electronic coherence may actually slow down the photosynthetic process.
\end{abstract}

\maketitle

\section{Introduction}\label{sec:introduction}

There has been much excitement in recent years about the observation of surprisingly long-lived quantum coherence (presence of non-zero off-diagonal elements in a quantum mechanical density matrix) in the excitation energy transfer in photosynthetic systems \cite{Engel2007,Ishizaki2009a,Panitchayangkoon2010,Ishizaki2010,Collini2010,Strumpfer2012,Chin2013}. Since being first observed, it has been suggested that this coherence may be important for efficient transfer of the electronic excitation.

Much of the study of this effect has focussed on the Fenna-Matthews-Olson (FMO) complex \cite{Fenna1974}, which is found in green sulfur bacteria such as \emph{C. tepidum} and \emph{P. aestuarii}, and funnels excitation energy, collected by a light-harvesting chlorosome, towards a photosynthetic reaction centre. The complex comprises bacteriochlorophyll (BChl) pigments bound to a protein, and is a trimer, with each monomer containing 8 BChl molecules (the existence of the eighth having been confirmed only relatively recently \cite{Tronrud2009,SchmidtamBusch2011}). Recently, the FMO complex has been the most popular system for both experimental and theoretical studies of energy transfer in photosynthesis, due to its small size and simplicity.

The 2-dimensional electronic spectroscopy experiments of Engel \emph{et al.} \cite{Engel2007}, carried out on the FMO complex, showed long-lived quantum beating at cryogenic temperature (77 K). A follow-up theoretical study by Ishizaki and Fleming \cite{Ishizaki2009a} at both 77 K and 300 K suggested that quantum coherent energy transfer might also be observed in the complex at 300 K. Since then, long-lived quantum beating has indeed been observed in 2D spectroscopy experiments of the FMO complex at room temperature \cite{Panitchayangkoon2010}. This electronic coherence has been the subject of scores of recent theoretical studies of energy transfer in photosynthesis (see  \cite{Pachon2012} for a recent review of these studies).

However, the question of whether or not quantum coherence is important for the energy transfer is controversial. In particular, there have been a number of recent papers that question the nature of this coherence \cite{Briggs2011,Miller2012,Leon-Montiel2013}, showing that classical theories can account for long-lived oscillations. Nonetheless, in much of the literature the incoherent F\"orster theory is dismissed as a predictive tool for the energy transfer dynamics due to the fact that it neglects coherence \cite{Cheng2009,Ishizaki2009}.

The existence of methods such as the Hierarchical Equations of Motion (HEOM), discussed in Section \ref{sec:theory}, allows the excitation energy transfer dynamics in the FMO complex to be simulated with numerically exact accuracy for a given Hamiltonian, meaning that the validity of approximate techniques can be tested. Comparing the predictions of F\"orster theory to those of the HEOM will allow us to appraise whether or not quantum coherence is truly important in the energy transfer in the FMO complex. We will consider transfer at 300 K, since this is more relevant to the biophysics of the complex, and will look at electronic energy transfer throughout the entire 24-site trimer.

We first run a fully numerically exact calculation for the canonical model of the 7-site FMO subsystem, and show that the approximation made by Ishizaki and Fleming in Eq. A2 of \cite{Ishizaki2009} leads to a noteworthy deviation from the exact dynamics. This means that various authors who have used the calculations from \cite{Ishizaki2009} as a benchmark to demonstrate the validity of their newly introduced methods (for example, \cite{Huo2011,Ritschel2011,Berkelbach2012,Seuss2014}) may not have been aware that their methods were in fact less or more accurate than it originally appeared based on the comparison with \cite{Ishizaki2009}.

Then we look at long-time dynamics. Most simulations of the FMO complex published in the literature only look at the first 1 ps \cite{Ishizaki2009a,Ritschel2011} since electronic coherence decays on a timescale of around 400 fs. However, after the decay of coherence, the populations decay to a steady state \cite{Huo2011,Zhu2011}, and it is illuminating to look at both short- and long-time dynamics, in order to test the performance of F\"orster theory.

The fluorescence lifetime of the FMO complex is on the order of 1 ns \cite{Louwe1997}, and in this work the timescales of excitation energy transfer are found to be about 15 ps. Since the energy transfer rate is thus more than 50 times faster than that of spontaneous emission, loss of population to the latter process can be neglected. The timescale of excitation trapping at the reaction centre has been reported to be on the order of picoseconds, based on the time taken for equilibrium to be reached \cite{Plenio2008a}. However, without treating the reaction centre explicitly, nothing can be said about the efficiency of excitation transfer in the two methods.

The remainder of this paper is set out as follows: Section \ref{sec:theory} introduces the model used to describe excitation energy transfer, as well as the exact HEOM and approximate F\"orster theory. Section \ref{sec:results} compares the energy transfer dynamics predicted by the coherent and incoherent theories, and Section \ref{sec:discussion} concludes.

\section{Theory}\label{sec:theory}

In the model we will use to describe the FMO complex \cite{Ishizaki2009a,Ritschel2011}, we treat each BChl molecule as having two electronic energy levels, and denote by $\ket{j}$ the state in which BChl $j$ is in its excited state and all others are in their ground states. Then, the so-called system Hamiltonian for $N$ BChl sites is:
\begin{subequations}
\begin{equation}
\hat{H}_{S} = \sum_{j=1}^{N}\hbar\omega_{j}\ket{j}\bra{j} + \sum_{i=1}^{N-1}\sum_{j=i+1}^{N}J_{ij}(\ket{i}\bra{j} + \ket{j}\bra{i}),
\end{equation}

\noindent with $\hbar\omega_{j}$ the energy of site $j$ and $J_{ij}$ the dipolar coupling between sites $i$ and $j$. Since we want to treat $\hat{H}_S$ as an open quantum system, each site is associated with a phonon bath consisting of the vibrations of the BChl molecule and the surrounding protein. Each bath has an infinite number of oscillators, and the bath Hamiltonian is:
\begin{equation}
\hat{H}_{B} = \sum_{j=1}^{N}\sum_{\kappa}\left(\frac{\hat{p}_{j\kappa}^{2}}{2m_{j\kappa}} + \frac{1}{2}m_{j\kappa}\omega_{j\kappa}^{2}\hat{q}_{j\kappa}^{2}\right),
\end{equation}

\noindent where $\hat{q}_{j\kappa}$ is the coordinate of the $\kappa^{\text{th}}$ phonon mode associated with site $j$, $\hat{p}_{j\kappa}$ is its momentum, $m_{j\kappa}$ its mass, and $\omega_{j\kappa}$ its angular frequency. Finally, there is a vibronic interaction between each site and its bath, given by the system-bath coupling Hamiltonian:
\begin{equation}
\hat{H}_{SB} = -\sum_{j=1}^{N}\hat{V}_{j}\hat{u}_{j}.
\end{equation}
\end{subequations}

\noindent Here,  $\hat{V}_{j} \equiv \ket{j}\bra{j}$, and $\hat{u}_{j} = \sum_{\kappa}g_{j\kappa}\hat{q}_{j\kappa}$, with $g_{j\kappa}$ describing the strength of the coupling of the $\kappa^{\text{th}}$ mode to the $j^{\text{th}}$ site. Then the total Hamiltonian for the FMO complex is given by:
\begin{equation}\label{eq:total_hamiltonian}
\hat{H} = \hat{H}_{S} + \hat{H}_{B} + \hat{H}_{SB}.
\end{equation}

The bath has an infinite number of degrees of freedom, but since we are not interested in the dynamics of the bath degrees of freedom, we can average over them. The reduced (system) density operator is defined as a trace over the  bath degrees of freedom in the total density operator $\hat{\rho}(t)$:
\begin{equation}\label{eq:rdo}
\hat{\rho}_{S}(t) = \text{tr}_{B}[\hat{\rho}(t)].
\end{equation}

The influence of the bath on the system can be fully characterized by the spectral density, $J_{j}(\omega)$ for the phonons coupled to site $j$:
\begin{equation}\label{eq:spectral_density}
J_{j}(\omega) = \sum_{\kappa}\frac{g_{j\kappa}^{2}}{2m_{j\kappa}\omega_{j\kappa}}\delta(\omega-\omega_{j\kappa}).
\end{equation}

In addition, the bath correlation function $\alpha_{j}(t)$ is defined such that $\alpha_{j}(t_1,t_2)=\alpha_{j}(t_1-t_2) = \left\langle \tilde{u}_{j}(t_1)\tilde{u}_{j}(t_2)\right\rangle_{\beta}$, with $\left<\dots\right>_{\beta}$ representing a thermal average over the bath degrees of freedom, and $\tilde{u}_{j}(t) = e^{\rm{i}\hat{H}_{B}t/\hbar}\hat{u}_{j}e^{-\rm{i}\hat{H}_{B}t/\hbar}$. This function is related to the spectral density by the expression:
\begin{equation}\label{eq:corrfunc_integral}
\alpha_{j}(t) = \int_{0}^{\infty}J_{j}(\omega)\left(\coth(\beta\hbar\omega/2)\cos(\omega t) - \text{i}\sin(\omega t)\right)d\omega.
\end{equation}

In the following, we write the correlation function as a sum of exponentials:
\begin{equation}\label{eq:alpha_exponentials}
\alpha_{j}(t) = \sum_{k}^{K}p_{jk}e^{-\gamma_{jk}t},
\end{equation}

\noindent with $p_{jk}$ and $\gamma_{jk}$ defined by the Pad\'{e} decomposition that was first introduced in \cite{Hu2010}, and explained in more detail in \cite{Hu2011}.

We now introduce the two methods used in this paper to calculate the reduced density operator dynamics for this system.

\subsection{Hierarchical Equations of Motion}

The HEOM were originally developed by Tanimura \cite{Tanimura1989,Tanimura1990}, and give a numerically exact method for calculating the dynamics of the reduced density operator $\rho(t)$, placing it at the bottom of a hierarchy of auxiliary density matrices (ADOs), denoted $\hat{\rho}_{\textbf{n}}(t)$. $\textbf{n}$ is an $N\times K$ matrix, with $N$ the number of baths and $K$ the number of exponential terms in Eq. \eqref{eq:alpha_exponentials}; an element of $\textbf{n}$ is denoted $n_{jk}$; all of these elements must be non-negative.

The $n^{\text{th}}$ level of the hierarchy consists of all ADOs for which:

\begin{equation}
\sum_{j=1}^{N}\sum_{k=1}^{K} n_{jk} = n,
\end{equation}

\noindent and the matrix with $\textbf{n} = \textbf{0}$ is the reduced density operator. In principle, the hierarchy has an infinite number of layers, but in practice a finite number of levels is sufficient for converged results. The equation of motion for an ADO is \cite{Ishizaki2005,Ishizaki2009}:
\begin{multline}\label{eq:heom}
\frac{\text{d}}{\text{d}t}\hat{\rho}_{\textbf{n}}(t) = -\frac{\text{i}}{\hbar}\left[\hat{H}_{S},\hat{\rho}_{\textbf{n}}(t)\right] - \sum_{j=1}^{N}\sum_{k=1}^{K}n_{jk}\gamma_{jk}\hat{\rho}_{\textbf{n}}(t) + \\
\text{i}\sum_{j=1}^{N}\sum_{k=1}^{K}\left(\left[\hat{V}_{j},\hat{\rho}_{\textbf{n}_{jk+}}(t)\right] + n_{jk}p_{jk}\hat{V}_{j}\hat{\rho}_{\textbf{n}_{jk-}}(t) + n_{jk}p_{jk}^{\ast}\hat{\rho}_{\textbf{n}_{jk-}}(t)\hat{V}_{j} \right),
\end{multline}

\noindent where $\hat{\rho}_{\textbf{n}_{jk\pm}}(t)$ is indexed by a matrix almost identical to that of $\hat{\rho}_{\textbf{n}}(t)$, but with the element $n_{jk}$ replaced by $n_{jk}\pm 1$. Thus, each ADO is coupled to operators in the level above and in the level below it in the hierarchy. If $n_{jk}=0$, then $\hat{\rho}_{\textbf{n}_{jk-}}(t) = 0$, and similarly, if $\hat{\rho}_{\textbf{n}}(t)$ is on the highest level of the hierarchy, then $\hat{\rho}_{\textbf{n}_{jk+}}(t) = 0$ for all $j$ and $k$.

Rather than solving this set of coupled differential equations using the traditional RK4 method, we used the fact that the equations are linear in the elements of the ADOs. If $\underline{\rho}(t)$ is a vector containing all of these elements, then the HEOM can be written in terms of a linear transformation $T$:
\begin{equation}\label{eq:linear}
T:\underline{\rho}(t)\rightarrow \frac{\text{d}}{\text{d}t}\underline{\rho}(t).
\end{equation}

Since $T$ is time-independent, Eqn. \eqref{eq:linear} has the formal solution:

\begin{equation}
\underline{\rho}(t + \Delta t) = \exp\left(T\Delta t\right)\underline{\rho}(t).
\end{equation}

\noindent Expanding this as a power series gives:

\begin{equation}\label{eq:power_series}
\underline{\rho}(t+\Delta t) = \sum_{l=0}^{L}\frac{(\Delta t)^{l}}{l!}T^{l}\underline{\rho}(t) + \mathcal{O}\left((\Delta t)^{L+1}\right),
\end{equation}

\noindent where $T^{l}\underline{\rho}(t)$ is short for the $l^{\text{th}}$ derivative of $\underline{\rho}(t)$, calculated using Eqn. \eqref{eq:heom}. This avoids calculating a matrix representation for $T$, which is potentially very large (though sparse). The algorithm we used to propagate $\underline{\rho}(t)$ through time used a second vector, $\underline{\rho}^{\prime}$, for storage:

\begin{enumerate}
\item Set $\underline{\rho}^{\prime} = \underline{\rho}(t)$.
\item Set $\underline{\rho}(t+\Delta t) = \underline{\rho}(t)$.
\item For $l = 1$ to $L$:
\begin{itemize}
\item Set $\underline{\rho}^{\prime} = \frac{\Delta t}{l}T\underline{\rho}^{\prime}$,
\item Set $\underline{\rho}(t+\Delta t) = \underline{\rho}(t+\Delta t) + \underline{\rho}^{\prime}$.
\end{itemize}
\end{enumerate}

For $L=4$, this method has the same order of convergence as the RK4 method, but the latter requires the storage of five vectors for each time-propagation step, whereas the method presented above requires storage of only one vector $\underline{\rho}^{\prime}$, which is a significant computational advantage. The parameter $L$ is also variable: a smaller $L$ requires a smaller timestep, but fewer calculations of the time derivative. We found that $L = 2$ was the best compromise, and gave the most efficient time-integration.

\subsection{F\"orster Theory}

F\"orster theory was originally formulated to describe resonant energy transfer between two electronic states \cite{Forster1948,Scholes2003}. It is a perturbative theory, which works well if the magnitudes of the dipolar interactions $|J_{ij}|$ between sites are sufficiently smaller than the transition energies $\hbar|\omega_{i}-\omega_{j}|$ or the reorganization energy $\lambda_{j} = \int_{0}^{\infty}\frac{J(\omega)}{\omega}d\omega$ \cite{Renger2009}. Since the transition energies are generally larger than the dipolar couplings in the FMO complex (in Eqn. \eqref{eq:adolphs_renger}, for example, the largest dipolar coupling, $H_{12}$, has magnitude 87.7 cm$^{-1}$, while $\hbar|\omega_{1}-\omega_{2}| = 120$ cm$^{-1}$), we might expect this to be a reasonable approximation. In F\"orster theory, only the diagonal matrix elements of the reduced density operator (the site populations) are evolved through time, while the off-diagonal elements (coherences) are set equal to zero, meaning that the theory is fully incoherent \cite{Yang2002}.

The theory involves a set of coupled rate equations for the site populations $\rho_{jj}(t) = \bra{j}\hat{\rho}_{S}(t)\ket{j}$:
\begin{equation}\label{eq:rate_equations}
\frac{\text{d}}{\text{d}t}\rho_{jj}(t) = \sum_{i=1}^{N}\sum_{j\ne i}^{N}\left( k_{i\rightarrow j}\rho_{ii}(t) - k_{j\rightarrow i}\rho_{jj}(t)\right),
\end{equation}

\noindent with the transfer rate $k_{i\rightarrow j}$ given by \cite{Yang2002}:
\begin{equation}\label{eq:forster_rate}
k_{i\rightarrow j} = 2|J_{ij}|^{2}\Re\left(\int_{0}^{\infty}F_{i}^{\ast}(t)A_{j}(t)dt\right),
\end{equation}

\noindent where $\Re$ denotes the real part. $F_{i}(t)$ and $A_{j}(t)$ are the fluorescence and absorption lineshape functions, respectively:
\begin{subequations}
\begin{equation}\label{eq:fluorescence}
F_{i}(t) = \exp\left(-\text{i}(\omega_{i}-\lambda_{i}/\hbar)t - g_{i}^{\ast}(t)\right),
\end{equation}
\begin{equation}\label{eq:absorption}
A_{j}(t) = \exp\left(-\text{i}(\omega_{j}+\lambda_{j}/\hbar)t - g_{j}(t)\right).
\end{equation}
\end{subequations}

The line-broadening function is $g_{j}(t) = \frac{1}{\hbar^{2}} \int_{0}^{t}dt_{1}\int_{0}^{t_{1}}dt_{2}\alpha_{j}(t_{2})$. As explained by Yang and Fleming \cite{Yang2002}, the fluorescence spectrum of site $i$ is the Fourier transform of $F_{i}(t)$, and the absorption spectrum of site $j$ is the Fourier transform of $A_{j}(t)$.

This represents a simple, easily implemented and computationally cheap theory for incoherent excitation energy transfer, and in Section \ref{sec:results}, the comparison between the two theories will give an idea of how well the incoherent theory can predict the energy transfer dynamics.

\section{Results}\label{sec:results}

In order to simulate the exact and approximate energy transfer dynamics for the 24-site FMO trimer and for the 7-site system, we first present the physical parameters which we will use; all of this data is found in the Appendix.

The 7-site model system Hamiltonian we use is that of Adolphs and Renger \cite{Adolphs2006}, which is popular in theoretical studies of this system \cite{Ishizaki2009a,Ritschel2011,Nalbach2011}. The parameters in this model are given in Eqn. \eqref{eq:adolphs_renger}.

For the 24-site calculations we follow the study by Ritschel \emph{et al.} \cite{Ritschel2011}, using the Hamiltonian given in Eqns. \eqref{eq:ritschel}, with two sets of site energies as in Eqn. \eqref{eq:site_energies}: those of Olbrich \emph{et al.} (OLB) \cite{Olbrich2011} and those of Schmidt am Busch \emph{et al.} (SAB) \cite{SchmidtamBusch2011}.

The spectral density chosen is the Lorentz-Drude/Debye spectral density:
\begin{equation}
J(\omega) = \frac{2}{\pi}\frac{\lambda\gamma\omega}{\gamma^{2} + \omega^{2}},
\end{equation}

\noindent with $\lambda = 35~\text{cm}^{-1}$ and $\gamma = 106.1767~\text{cm}^{-1}$ (with the same spectral density used for every site). These parameters are used in the most popular benchmark study of energy transfer in the FMO \cite{Ishizaki2009a}, which used this form for the spectral density because it had been used successfully in the analyses of several experiments, in particular Refs.  \cite{Read2008,Zigmantas2006,Read2007}. Using the Pad\'e approximant to the Bose-Einstein function in the definition of $\alpha_{j}(t)$ \cite{Hu2011} represents $\alpha_j(t)$ in the form of Eqn. \eqref{eq:alpha_exponentials}. Convergence was obtained with two terms, with the values of the $p_{jk}$ and the $\gamma_{jk}$ given by Eqn. \eqref{eq:pade_alphas} in the Appendix.

Since BChl molecules 1, 6 and 8 are closest to the chlorosome, it is these that are most likely to be excited initially, and so the simulations are started with excitation on one of these three sites. In each of the results shown, the populations of a number of sites are omitted, as they remain small throughout the simulation.

\begin{figure*}[!ht]
\includegraphics[width=1\textwidth]{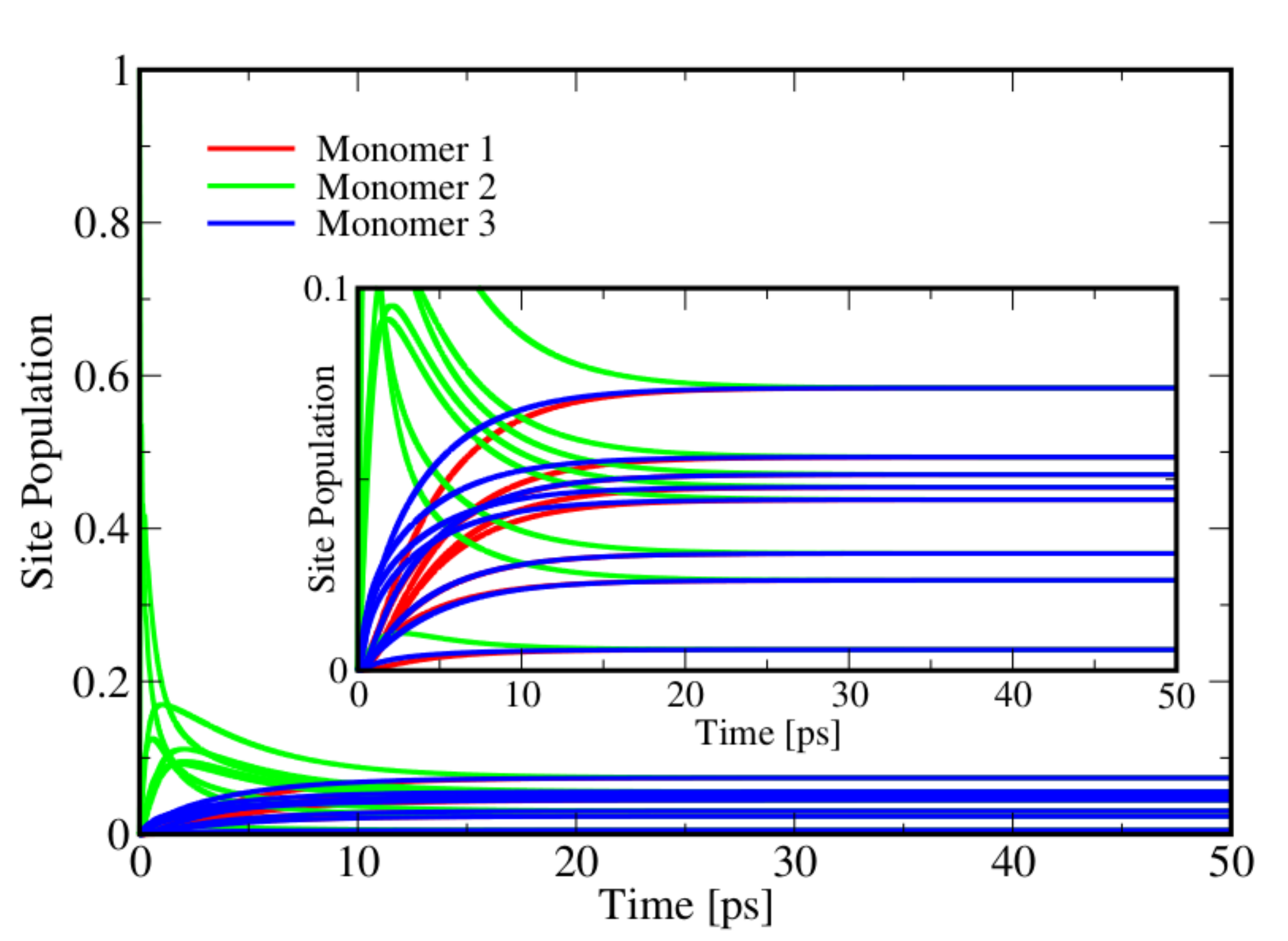}
\caption{\label{fig:steadystates}\small{Numerically exact dynamics for the full 24-site trimer up to the steady state. Each colour represents a particular monomer, as indicated in the legend.}}
\end{figure*}

We begin by showing our numerically exact calculations of the full 24-site FMO trimer, in Fig. \ref{fig:steadystates}. It is interesting to note that regardless of which monomer was initially excited, all monomers are equally populated at equilibrium, and the population distribution is the same in each monomer. This is a consequence of the threefold rotational symmetry of the FMO complex,  in which we see that for a given site, the population at long times is the same on each monomer regardless of which one was initially excited.

\subsection{7-Site System}

\begin{figure*}[!ht]
\includegraphics[width=14cm]{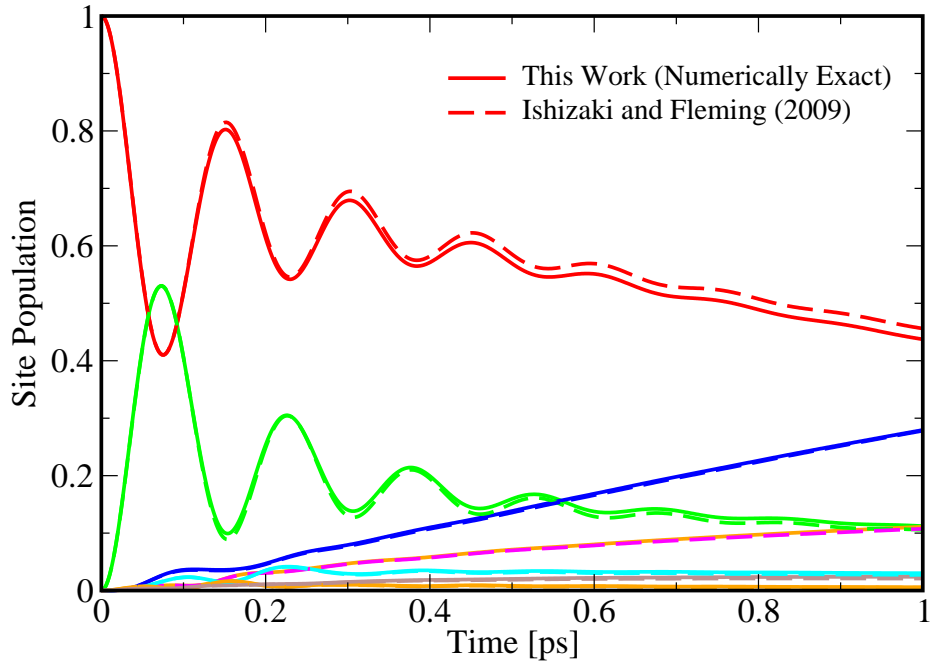}
\caption{\label{fig:ishizakifleming}\small{Comparison of HEOM results with full bath correlation function (solid lines) and those of Ref. \cite{Ishizaki2009a} (dashed), with bath correlation function approximated by one exponential and a delta-function.}}
\end{figure*}

Fig. \ref{fig:ishizakifleming} compares the population dynamics at 77 K after convergence is achieved with respect to the number of exponentials used to represent $\alpha(t)$ (3 exponentials were used here), to the population dynamics presented by Ishizaki and Fleming in \cite{Ishizaki2009a}; which were for the same parameters, but with $\alpha(t)$ truncated as described by the approximation shown in Eq. A2 of their paper \cite{Ishizaki2009a}. The results we present here are thus the first numerically exact calculations for energy transfer in the FMO complex. The difference is expected to be even larger at lower temperatures where more exponentials are required. 

\begin{figure*}[!ht]
\includegraphics[width=16cm]{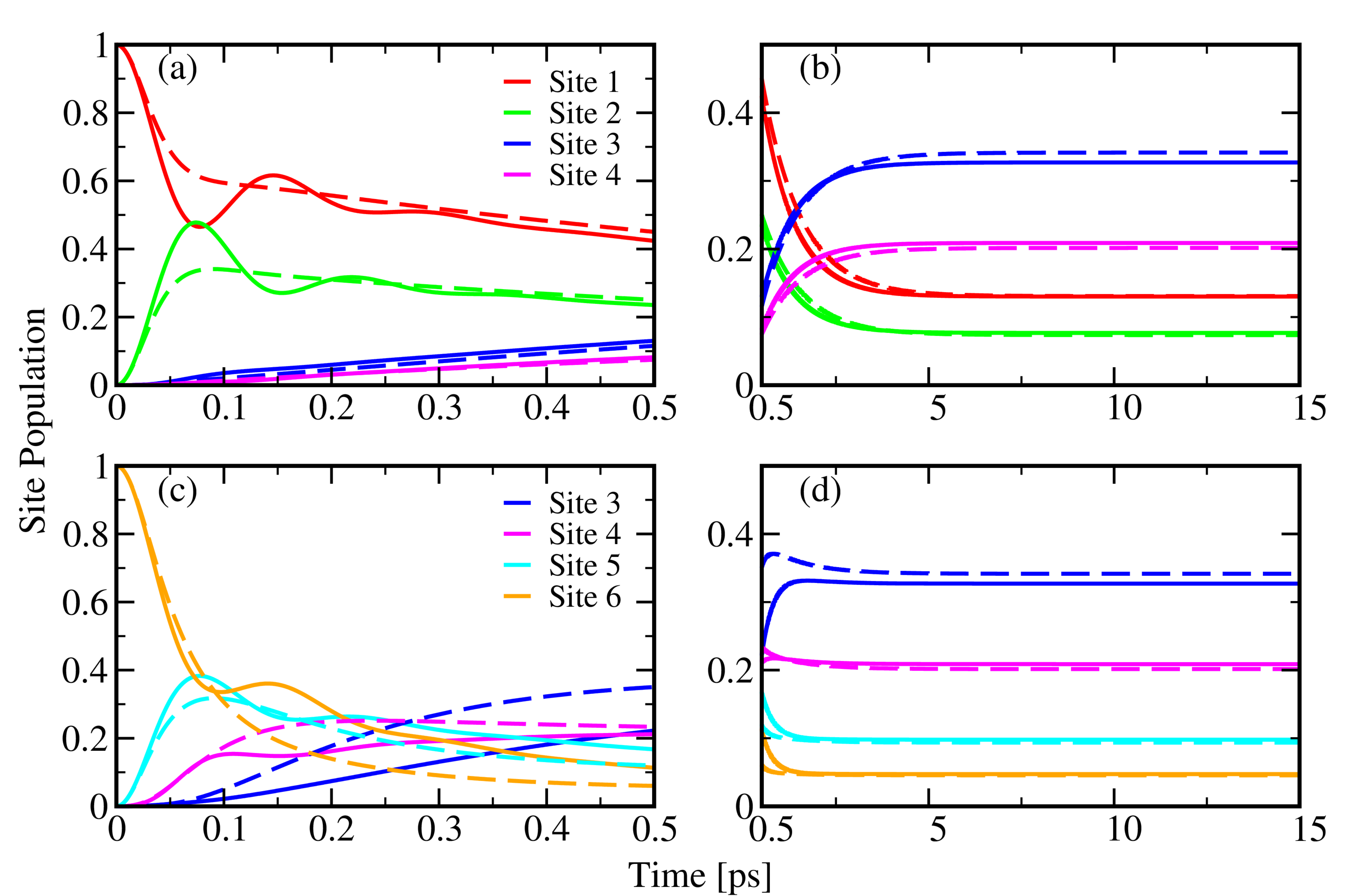}
\caption{\label{fig:7site}\small{Comparison of HEOM (solid lines) and F\"orster theory (dashed lines) dynamics for the 7-site subsystem. Panels (a) and (b) correspond to initial excitation on site 1, and (c) and (d) correspond to initial excitation on site 6. The left-hand column focuses on short-time dynamics, and the right-hand column shows the longer-time dynamics.}}
\end{figure*}

In Fig. \ref{fig:7site}, we compare our fully converged dynamics of HEOM, to incoherent F\"orster theory for the 7-site model, with initial excitation on sites 1 and 6, showing both the short-time and the longer-time dynamics. The HEOM results were obtained with 2 exponential terms in Eqn. \eqref{eq:alpha_exponentials} (i.e., $K=2$; the numerical values for these terms are given by Eqn. \eqref{eq:pade_alphas} in the Appendix) and 5 layers of the hierarchy, which we found to be sufficient to fully converge the dynamics. The F\"orster theory calculations used the same bath correlation function, with $K=2$.

Looking at the results of the F\"orster theory, we note that while perfect agreement with the exact results is impossible - after all, this method can never capture the electronic coherences - it is still surprisingly good: for the upper panels of Fig. \ref{fig:7site}, in which site 1 is initially excited, the incoherent theory gives a very good prediction of the transfer rates, steady-state populations, and timescale on which the steady state is reached.

The lower panels of Fig. \ref{fig:7site}, where site 6 is excited initially, show a somewhat less impressive comparison between the two methods: the steady-state populations are, naturally, independent of which site was populated at the beginning of the calculation, and so are matched just as well as in the upper panels, and the timescale on which the steady-state was reached is also captured well, but the transfer rates at short times are predicted quite poorly by F\"orster theory. However, the rates are over-predicted, which is unexpected: site 3, which is nearest to the reaction centre, is predicted to have a greater population before about 2.5 ps in the incoherent theory than is seen in the exact results, which suggests that if transfer to the reaction centre were to occur on this timescale, then incoherent transfer might be faster than coherent transfer.

\subsection{24-Site System}

Almost all of the study in the literature on energy transfer in the FMO complex as an open quantum system has only treated a subsystem of the full complex \cite{Ishizaki2009a,Huo2010,Nalbach2011,Ritschel2011a,Moix2011}. The only study of the full 24-site model to date was that of Ritschel \emph{et al.} \cite{Ritschel2011}, which used an approximate quantum master equation to simulate the dynamics up to 1 ps. However, one very interesting aspect that is new in the study of the full trimer is the emergence of a second timescale in the transfer dynamics. While the time-scale for intra-monomer energy transfer is expected to be on the order of $\left|\hbar/J_{ij}\right| \sim 400\text{~fs}$, with $J_{ij}$ an off-diagonal element taken from Eqn. \eqref{eq:intramonomer} (specifically, between one of sites 1, 6 or 8 and its most strongly-coupled neighbour), the time-scale for inter-monomer transfer will be on the order of $\left|\hbar/V_{ij}\right| \sim 20\text{~ps}$, where $V_{ij}$ is an element from Eqn. \eqref{eq:intermonomer}. Therefore, to get an idea of the dynamics of the 24-site model all the way to equilibrium, we have calculated the dynamics up to 20 ps.

\begin{figure*}[htp]
\includegraphics[width=16cm]{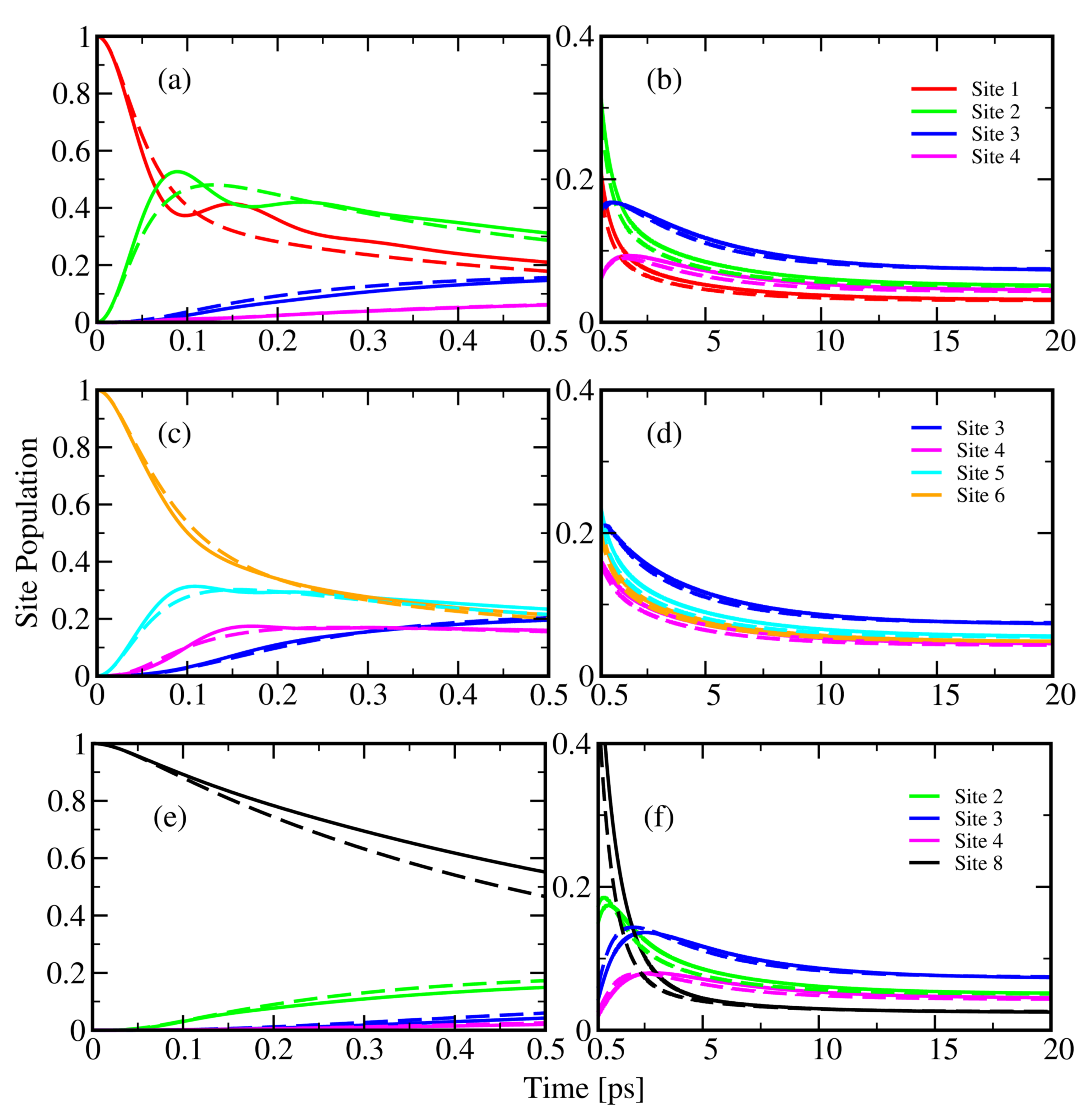}
\caption{\small{Comparison of HEOM (solid lines) and F\"orster theory (dashed lines) dynamics for the monomer with the initial excitation, with the full 24-site system considered, with OLB site energies. Initial excitation on Site 1 (a)-(b), Site 6 (c)-(d), Site 8 (e)-(f). The left-hand column shows short-time dynamics and the right-hand column, long-time dynamics.}}\label{fig:OLB}
\end{figure*}

\begin{figure*}[htp]
\includegraphics[width=17cm]{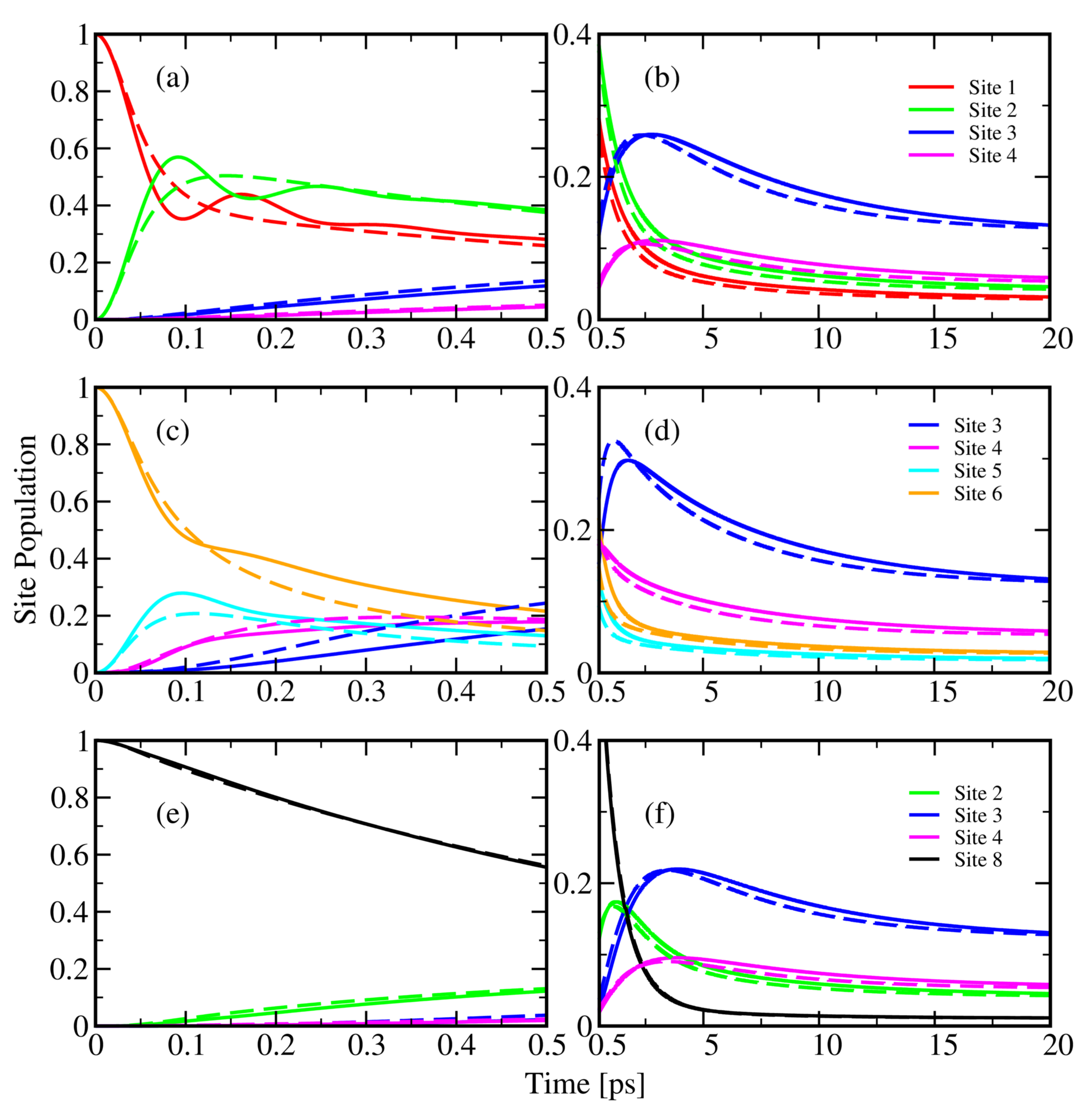}
\caption{\small{Comparison of HEOM (solid lines) and F\"orster theory (dashed lines) dynamics for the monomer with the initial excitation, with the full 24-site system considered, with SAB site energies. Initial excitation on  Site 1 (a)-(b), Site 6 (c)-(d), Site 8 (e)-(f). The left-hand column shows short-time dynamics and the right-hand column, long-time dynamics.}}\label{fig:SAB}
\end{figure*}

Fig. \ref{fig:OLB} compares the dynamics of HEOM and F\"orster theory for the 24-site model using the OLB site energies, for initial excitation on sites 1, 6 and 8 and for both short and longer timescales. Fig. \ref{fig:SAB} repeats this comparison for the 24-site model using the SAB site energies. In each case, the bath correlation function contained two terms, and for the HEOM results, 4 levels of the hierarchy were used.


As in the 7-site system, the results of the HEOM calculations show coherences between sites 1 and 2 at short timescales for both site energies, but little coherence is observed
 between sites 5 and 6. When site 8 is initially excited, the transfer is incoherent; similar dynamics were reported at 77 K in Ref. \cite{Ritschel2011}, and at 300 K in the 8-site monomer in Ref. \cite{Moix2011}. In both Hamiltonians, despite the fact that site 8 is most strongly coupled to site 1, $J_{18} < \hbar\left|\omega_{1}-\omega_{8}\right|$. The difference is much smaller in the OLB site energies than in the SAB energies, suggesting that in the latter, coherence between these two sites will be minimal, though coherence may be seen in the former. In addition, the strong coupling between sites 1 and 2 means that excitation will flow to the 2$^{\text{nd}}$ site before coherence could develop between sites 8 and 1\cite{Ritschel2011}. Whether or not coherence occurs at short timescales, it is quenched by around 400 fs, and as with the 7-site system the dynamics then enters an exponential decay-like regime, finally reaching a steady state. 

Once again, the predictions of F\"orster theory agree quite well with the exact dynamics: the general shapes of the population decays are followed with good accuracy, and both the timescales on which the steady state is reached, and the populations at this state, are captured -- generally with better accuracy than for the 7-site system. The middle row in Fig. \ref{fig:SAB}, in which the SAB Hamiltonian is used and site 6 is initially excited, shows relatively poor agreement, as does the bottom panel in Fig. \ref{fig:OLB}, in which the OLB Hamiltonian is used and site 8 is initially excited -- but once again, the transfer rates at short times are actually over-predicted by F\"orster theory. We have have observed the same wherever the incoherent theory noticeably fails to match the exact results.

This failure of F\"orster theory, even in a case where the population dynamics appear at first sight to be incoherent, can be compared with the bottom row of Fig. \ref{fig:SAB}, in which the two theories are matched very well at short times. As described above, some coherence is possible between sites 8 and 1 with the OLB energies, and this could be the reason for the failure of the incoherent theory. However, the performance of F\"orster theory is overall better for the full 24-site trimer than for the 7-site system.

\section{Discussion}\label{sec:discussion}

In this paper, we have presented a new method for solving the hierarchical equations of motion, and have used these to perform the first numerically exact calculation for the full 24-site FMO trimer and the 7-site subsystem of a monomer. We have also compared these dynamics to the approximate, perturbative F\"orster theory. The real upshot of our results is that, despite missing short-time oscillations, F\"orster theory provides quite a good description of energy transfer for the 3 popular models we have considered. This is, perhaps, unsurprising, because the applicability of the theory depends on the inter-site coupling being smaller than the transition energy between sites, which is often the case here. In a recent review, Ishizaki \emph{et al.} showed that, for a representative 2-site system, the energy transfer rate predicted by F\"orster theory should be similar to the exact transfer rate (Fig. 7 of \cite{Ishizaki2010}) for a reorganization energy of $\lambda = 35\text{~cm}^{-1}$. This is in accord with the results we have observed here for the 7- and 24-site systems.

The fact that, when the two theories give conflicting results at short times, F\"orster theory often tends to over-predict excitation transfer rates, is interesting: it has previously been assumed that electronic coherence is harnessed to provide the efficient transfer observed, but we have seen that an incoherent theory can predict faster transfer. However, while studying the 24-site system is an improvement on smaller subsystems, the trapping of excitation at the reaction centre is still omitted: this point is crucial, as without a better knowledge of the timescale of this trapping (which would require extending the Hamiltonian to include the reaction centre), we cannot single out which method gives the most efficient transfer.

However, assuming that the rate of excitation trapping is on the order of picoseconds, we notice that it is on this timescale that the two theories begin to agree: even if the exact, coherent, energy transfer was more efficient, there would be only a modest difference in efficiencies between the two.

F\"orster theory is valid in the same limits in which coherence in the density matrix dynamics is expected to be unimportant: the coupling between BChl sites is smaller than the transition energy between them. This means that, despite its simplicity, its reasonable predictions of transfer rates and long-time behaviour greatly support our conclusion that the FMO complex does not seem to utilize coherence to provide efficient transfer after all.

\section*{Acknowledgements}

The authors would like to thank David Manolopoulos for several interesting discussions, and Dmitri Iouchtchenko for careful proofreading of the manuscript. D.M.W. would like to thank the Clarendon Fund and St. Edmund Hall for financial support. N.S.D. thanks the Clarendon Fund, NSERC/CRSNG of/du Canada, and JSPS for financial support; and Yoshitaka Tanimura of Kyoto University for his much appreciated hosting in Kyoto.

\appendix

\section*{Appendix}

\subsection*{7-Site Hamiltonian}

The 7-site Hamiltonian of Adolphs and Renger is given by (energies in cm$^{-1}$)\cite{Adolphs2006}:
\begin{equation}\label{eq:adolphs_renger}
\hat{H}_{S} = 
\begin{pmatrix}
410 & -87.7 & 5.5 & -5.9 & 6.7 & -13.7 & -9.9 \\
    & 530   & 30.8& 8.2  & 0.7 & 11.8  & 4.3  \\
    &       & 210 & -53.5& -2.2& -9.6  & 6.0  \\
    &       &     & 320  & -70.7&-17.0 & -63.3\\
    &       &     &      & 480  & 81.1 & -1.3 \\
    &       &     &      &      & 630  & 39.7 \\
    &       &     &      &      &      & 440
\end{pmatrix}.
\end{equation}

\subsection*{24-Site Hamiltonian}

The 24-site Hamiltonian has the form:
\begin{subequations}\label{eq:ritschel}
\begin{equation}
\hat{H}_{S} = 
\begin{pmatrix}
\hat{h}_{A} & \hat{h}_{B} & \hat{h}_{B}^{\dagger} \\
\hat{h}_{B}^{\dagger} & \hat{h}_{A} & \hat{h}_{B} \\
\hat{h}_{B} & \hat{h}_{B}^{\dagger} & \hat{h}_{A}
\end{pmatrix},
\end{equation}

\noindent with the inter-site couplings (off-diagonal elements of $\hat{h}_{A}$) given by:
\begin{equation}\label{eq:intramonomer}
\hat{h}_{A} = 
\begin{pmatrix}
 \hspace{12pt}& -80.3	& 3.5 	& -4.0 	& 4.5 	& -10.2 & -4.9 	& 21.0 	\\
	&		& 23.5	& 6.7	& 0.5	& 7.5	& 1.5	& 3.3 	\\
	&		&		& -49.8	& -1.5	& -6.5	& 1.2	& 0.7	\\
	&		&		&		& -63.4	& -13.3	& -42.2	& -1.2	\\
	&		&		&		&		& 55.8	& 4.7	& 2.8	\\
	&		&		&		&		&		& 33.0	& -7.3	\\
	&		&		&		&		&		&		& -8.7  \\
	&		&		&		&		&		&		&
\end{pmatrix}.
\end{equation}

There are two sets of site energies (diagonal elements of $\hat{h}_{A}$), those of Olbrich \emph{et al.}\cite{Olbrich2011} (OLB) and those of Schmidt am Busch \emph{et al.}\cite{SchmidtamBusch2011} (SAB):
\begin{equation}\label{eq:site_energies}
\begin{matrix}
Site & \vline & 1 & 2 & 3 & 4 & 5 & 6 & 7 & 8 \\ \hline
\text{OLB} & \vline & 186 & 81 & 0 & 113 & 65 & 89 & 492 & 218 \\
\text{SAB} & \vline & 310 & 230 & 0 & 180 & 405 & 320 & 270 & 505 .
\end{matrix}
\end{equation}

Finally, the inter-site couplings $\hat{h}_{B}$ are given by:
\begin{equation}\label{eq:intermonomer}
\hat{h}_{B} = 
\begin{pmatrix}
1.0	& 3.0 & -0.6 & 0.7 & 2.3 & 1.5 & 0.9 & 0.1 \\
1.5 & -0.4 & -2.5 & -1.5 & 7.4 & 5.2 & 1.5 & 0.7 \\
1.4 & 0.1 & -2.7 & 5.7 & 4.6 & 2.3 & 4.0 & 0.8 \\
0.3 & 0.5 & 0.7 & 1.9 & -0.6 & -0.4 & 1.9 & -0.8 \\
0.7 & 0.9 & 1.1 & -0.1 & 1.8 & 0.1 & -0.7 & 1.3 \\
0.1 & 0.7 & 0.8 & 1.4 & -1.4 & -1.5 & 1.6 & -1.0 \\
0.3 & 0.2 & -0.7 & 4.8 & -1.6 & 0.1 & 5.7 & -2.3 \\
0.1 & 0.6 & 1.5 & -1.1 & 4.0 & -3.1 & -5.2 & 3.6
\end{pmatrix}.
\end{equation}
\end{subequations}

\subsection*{Bath Correlation Function}

The integral in Eqn. \eqref{eq:corrfunc_integral} is evaluated for the Drude-Lorentz/Debye spectral density using the [0/1] Pad\'e series for the Bose-Einstein function, $f_{\rm{Bose}}(\beta\hbar\omega) = (1-e^{-\beta\hbar\omega})^{-1}$ \cite{Hu2010,Hu2011}, giving two exponential terms in Eqn. \eqref{eq:alpha_exponentials}, which was sufficient for converged dynamics at 300 K. The coefficients $p_{jk}$ and $\gamma_{jk}$ in this equation are:
\begin{equation}\label{eq:pade_alphas}
\begin{matrix}
k & \vline & p_{jk} ~ / ~ \text{cm}^{-2} & \vline & \gamma_{jk} ~ / ~ \text{cm}^{-1} \\ \hline
1 & \vline & 14278.9908 - 3716.1862 \text{i} & \vline & 106.1767 \\
2 & \vline & 4818.3993 & \vline & 1615.1170 .
\end{matrix}
\end{equation}

%


\begin{thebibliography}{99}
\bibliographystyle{unsrt}
\bibitem{Engel2007}
Engel,~G.~S.; Calhoun,~T.~R.; Read,~E.~L.; Ahn,~T.-K.; Man\v{c}al,~T.; Cheng,~Y.-C.; Blankenship,~R.~E.; Fleming,~G.~R. \emph{Nature} \textbf{2007}, 446, 782--6
\bibitem{Ishizaki2009a}
Ishizaki,~A.; Fleming,~G.~R. \emph{Proc. Natl. Acad. Sci. USA} \textbf{2009}, 106, 17255--60
\bibitem{Panitchayangkoon2010}
Panitchayangkoon,~G.; Hayes,~D.; Fransted,~K.~A.; Caram,~J.~R.; Harel,~E.; Wen,~J.; Blankenship,~R.~E.; Engel,~G.~S. \emph{Proc. Natl. Acad. Sci. USA} \textbf{2010}, 107, 12766--70
\bibitem{Ishizaki2010}
Ishizaki,~A.; Calhoun,~T.~R.; Schlau-Cohen,~G.~S.; Fleming,~G.~R. \emph{Phys. Chem. Chem. Phys.} \textbf{2010}, 12, 7319--37
\bibitem{Collini2010}
Collini,~E.; Wong,~C.~Y.; Wilk,~K.~E.; Curmi,~P. M.~G.; Brumer,~P.; Scholes,~G.~D. \emph{Nature} \textbf{2010}, 463, 644--647
\bibitem{Strumpfer2012}
Str\"{u}mpfer,~J.; Sener,~M.; Schulten,~K. \emph{J. Phys. Chem. Lett.} \textbf{2012}, 3, 536--542
\bibitem{Chin2013}
Chin,~A.~W.; Prior,~J.; Rosenbach,~R.; Caycedo-Soler,~F.; Huelga,~S.~F.; Plenio,~M.~B. \emph{Nature Phys.} \textbf{2013}, 9, 113--118
\bibitem{Fenna1974}
Fenna,~R.~E.; Matthews,~B.; Olson,~J.; Shaw,~E. \emph{J. Mol. Biol.} \textbf{1974}, 84, 231--240
\bibitem{Tronrud2009}
Tronrud,~D.~E.; Wen,~J.; Gay,~L.; Blankenship,~R.~E. \emph{Photosynth. Res.} \textbf{2009}, 100, 79--87
\bibitem{SchmidtamBusch2011}
{Schmidt am Busch},~M.; Müh,~F.; {El-Amine Madjet},~M.; Renger,~T. \emph{J. Phys. Chem. Lett.} \textbf{2011}, 2, 93--98
\bibitem{Pachon2012}
Pach\'{o}n,~L.~A.; Brumer,~P. \emph{Phys. Chem. Chem. Phys.} \textbf{2012}, 14, 10094--108
\bibitem{Briggs2011}
Briggs,~J.; Eisfeld,~A. \emph{Phys. Rev. E} \textbf{2011}, 83
\bibitem{Miller2012}
Miller,~W.~H. \emph{J. Chem. Phys.} \textbf{2012}, 136, 210901
\bibitem{Leon-Montiel2013}
Le\'{o}n-Montiel,~R. d.~J.; Torres,~J.~P. \emph{Phys. Rev. Lett.} \textbf{2013}, 110, 218101
\bibitem{Cheng2009}
Cheng,~Y.-C.; Fleming,~G.~R. \emph{Annu. Rev. Phys. Chem.} \textbf{2009}, 60, 241--62
\bibitem{Ishizaki2009}
Ishizaki,~A.; Fleming,~G.~R. \emph{J. Chem. Phys.} \textbf{2009}, 130, 234111
\bibitem{Ritschel2011}
Ritschel,~G.; Roden,~J.; Strunz,~W.~T.; Aspuru-Guzik,~A.; Eisfeld,~A. \emph{J. Phys. Chem. Lett.} \textbf{2011}, 2, 2912--2917
\bibitem{Huo2011}
Huo,~P.; Coker,~D.~F. \emph{J. Chem. Phys.} \textbf{2011}, 135, 201101
\bibitem{Zhu2011}
Zhu,~J.; Kais,~S.; Rebentrost,~P.; Aspuru-Guzik,~A. \emph{J. Phys. Chem. B} \textbf{2011}, 115, 1531--7
\bibitem{Louwe1997}
Louwe,~R. J.~W.; Aartsma,~T.~J. \emph{J. Phys. Chem. B} \textbf{1997}, 101, 7221--7226
\bibitem{Plenio2008a}
Plenio,~M.~B.; Huelga,~S.~F. \emph{New J. Phys.} \textbf{2008}, 10, 113019
\bibitem{Hu2010}
Hu,~J.; Xu,~R.-X.; Yan,~Y. \emph{J. Chem. Phys.} \textbf{2010}, 133, 101106
\bibitem{Hu2011}
Hu,~J.; Luo,~M.; Jiang,~F.; Xu,~R.-X.; Yan,~Y. \emph{J. Chem. Phys.} \textbf{2011}, 134, 244106
\bibitem{Tanimura1989}
Tanimura,~Y.; Kubo,~R. \emph{J. Phys. Soc. Jpn.} \textbf{1989}, 58, 101--114
\bibitem{Tanimura1990}
Tanimura,~Y. \emph{Phys. Rev. A} \textbf{1990}, 41, 6676--6687
\bibitem{Ishizaki2005}
Ishizaki,~A.; Tanimura,~Y. \emph{J. Phys. Soc. Jpn.} \textbf{2005}, 74, 3131--3134
\bibitem{Forster1948}
F\"{o}rster,~T. \emph{Ann. Phys.} \textbf{1948}, 437, 55--75
\bibitem{Scholes2003}
Scholes,~G.~D. \emph{Annu. Rev. Phys. Chem.} \textbf{2003}, 54, 57--87
\bibitem{Renger2009}
Renger,~T. \emph{Photosynth. Res.} \textbf{2009}, 102, 471--85
\bibitem{Yang2002}
Yang,~M.; Fleming,~G.~R. \emph{Chem. Phys.} \textbf{2002}, 282, 163--180
\bibitem{Adolphs2006}
Adolphs,~J.; Renger,~T. \emph{Biophys. J.} \textbf{2006}, 91, 2778--97
\bibitem{Nalbach2011}
Nalbach,~P.; Braun,~D.; Thorwart,~M. \emph{Phys. Rev. E} \textbf{2011}, 84, 7
\bibitem{Olbrich2011}
Olbrich,~C.; Str\"{u}mpfer,~J.; Schulten,~K.; Kleinekath\"{o}fer,~U. \emph{J. Phys. Chem. Lett.} \textbf{2011}, \emph{2011}, 1771--1776
\bibitem{Read2008}
Read,~E.~L.; Schlau-Cohen,~G.~S.; Engel,~G.~S.; Wen,~J.; Blankenship,~R.~E.; Fleming,~G.~R. \emph{Biophys. J.} \textbf{2008}, 95, 847--56
\bibitem{Zigmantas2006}
Zigmantas,~D.; Read,~E.~L.; Man\v{c}al,~T.; Brixner,~T.; Gardiner,~A.~T.; Cogdell,~R.~J.; Fleming,~G.~R. \emph{Proc. Natl. Acad. Sci. USA} \textbf{2006}, 103, 12672--7
\bibitem{Read2007}
Read,~E.~L.; Engel,~G.~S.; Calhoun,~T.~R.; Man\v{c}al,~T.; Ahn,~T.~K.; Blankenship,~R.~E.; Fleming,~G.~R. \emph{Proc. Natl. Acad. Sci. USA} \textbf{2007}, 104, 14203--8
\bibitem{Huo2010}
Huo,~P.; Coker,~D.~F. \emph{J. Chem. Phys.} \textbf{2010}, 133, 184108
\bibitem{Ritschel2011a}
Ritschel,~G.; Roden,~J.; Strunz,~W.~T.; Eisfeld,~A. \emph{New J. Phys.} \textbf{2011}, 13, 113034
\bibitem{Moix2011}
Moix,~J.; Wu,~J.; Huo,~P.; Coker,~D.; Cao,~J. \emph{J. Phys. Chem. Lett.} \textbf{2011}, 2, 3045--3052
\bibitem{Berkelbach2012}
Berkelnach,~T.; Markland,~T.; Reichman,~D. \emph{J.Chem. Phys.} \textbf{2012}, 136, 084104
\bibitem{Seuss2014}
Seuss,~D.; Eisfeld,~A.; Strunz,~W. \emph{Phys. Rev. Lett.} \textbf{2014}, 113, 150403
\end{thebibliography}
\end{document}